\documentclass{PoS}

\title{Nucleosynthesis and chemical evolution of intermediate-mass stars: results from planetary nebulae}

\ShortTitle{Results from planetary nebulae}

\author{\speaker{Walter J. Maciel}\thanks{Work partially supported by FAPESP/CNPq.}\\
        University of S\~ao Paulo\\
        E-mail: \email{maciel@astro.iag.usp.br}}

\author{Roberto D. D. Costa\\
        University of S\~ao Paulo\\
        E-mail: \email{roberto@astro.iag.usp.br}}

\author{Thais E. P. Idiart\\
        University of S\~ao Paulo\\
        E-mail: \email{thais@astro.iag.usp.br}}

\abstract{Planetary nebulae (PN) are an excellent laboratory to investigate the nucleosynthesis 
and chemical evolution of intermediate mass stars. In these objects accurate abundances can be obtained 
for several chemical elements that are manufactured or contaminated by the PN progenitor stars, such as He, 
N, C, and also elements that were originally produced by more massive stars of previous generations, 
namely  O, Ne, Ar, and S. Some of these elements are difficult to study in stars, so that PN can be used in 
order to complement results obtained from stellar data.

In the past few years, we have obtained a large sample of PN with accurately derived abundances, including 
objects of different populations, namely the solar neighbourhood, the galactic disk and anticentre, the 
galactic bulge and the Magellanic Clouds. 

In this work, we present the results of our recent analysis of the chemical abundances of He, O, N, S, Ar 
and Ne in galactic and Magellanic Cloud PN. Average abundances and abundance distributions of all 
elements are determined, as well as distance-independent correlations. These correlations are 
particularly important, as they can be directly compared with the predictions of recent theoretical 
evolutionary models for intermediate mass stars.}

\FullConference{11th Symposium on Nuclei in the Cosmos\\
		19-23 July, 2010 \\
		Heidelberg, Germany.}

\begin{document}

\section{Introduction}
Planetary nebulae (PN) are an excellent laboratory to investigate the nucleosynthesis and chemical 
evolution of intermediate mass stars. Accurate abundances can be obtained for several chemical elements, 
including (i) those elements that are manufactured by the PN intermediate-mass progenitor stars (He, N, C), 
and (ii) also elements that were originally produced by more massive stars of previous generations 
(O, Ne, Ar, S). The abundances of the first class of elements  measured in PN include the original content 
previous to the formation of the progenitor stars and the contamination effects during the nuclear processes 
in these objects. As a consequence, PN can be used to study the nucleosynthetic processes in intermediate 
mass stars. On the other hand, elements such as O, Ne, etc. reveal the interstellar abundances at the time 
and place the progenitor stars were formed, so that the determination of their chemical abundances 
produces important observational constraints to the chemical evolution models for the galaxies hosting 
these objects. 

In the past few years, we have obtained a large sample of PN of different galactic populations with accurately 
derived abundances (cf. Maciel et al. \cite{mci09}, \cite{mci10}, and references therein). In this work we 
present average abundances and abundance distributions of several elements, as well as distance-independent 
abundance correlations that  can be directly compared with the predictions of recent theoretical evolutionary 
models for intermediate mass stars.

\section{The sample}

For the Milky Way, we have considered three different samples: (i) our own IAG/USP data (cf. Maciel et al. 
\cite{mci09}, \cite{mci10}), which includes 84 disk PN and 188 bulge PN, supplemented by 84 bulge nebulae 
from Stasi\'nska et al. \cite{stasinska}; (ii) Sample A, which includes 234 PN in the Milky Way disk (Maciel 
et al. \cite{mcu}, Maciel and Costa \cite{mc10}), and (iii) Sample B, which is a larger sample which 
includes literature data, presently containing 372 nebulae. For the Magellanic Clouds, we have again 
considered our own data (Maciel et al. \cite{mci09}), including 45 nebulae for the SMC and 23 objects in
the LMC, supplemented by data from Stasi\'nska et al. \cite{stasinska}, with 48 PN in the SMC and 106 in 
the LMC,  and Leisy and Dennefeld  \cite{leisy}, with 36 PN in the SMC and 120 in the LMC. As discussed 
by Maciel et al. \cite{mci09}, the merged sample maintains the homogeneity of the individual samples, 
in view of the similar methods employed, so that a more comprehensive sample can be obtained.

\section{Average abundances and abundance distributions}

We have determined average abundances of all studied elements both in the Galaxy and in the Magellanic 
Clouds. The He abundances are similar in all samples within the average uncertainties, and are as follows: 
He/H $\simeq$ 0.115, 0.105, and 0.095 for the Milky Way, the LMC and the SMC, respectively. Typical 
uncertainties in the He abundances are approximately 0.020 to 0.030, and the abundances range from 0.05 
to 0.18 (Maciel et al. \cite{mci10}). The O/H abundances, which are usually the best determined of all 
heavy elements considered here, also show a good agreement among all different samples for a given system. 
In all cases the LMC is richer than the SMC, as expected, and the average metallicity differences are usually 
in the range 0.3 to 0.5 dex, which is consistent with the metallicities given by Stanghellini 
\cite{stanghellini}, namely Z = 0.004 and Z = 0.008 for the SMC and LMC, respectively. The same pattern is 
observed as the Milky Way data are included, as expected from the larger metallicity of the Galaxy relative 
to the Magellanic Clouds. The Ar/H and Ne/H ratios show a behaviour similar to O/H in all cases. The sulfur 
abundances seem to be less reliable, since larger standard deviations are obtained and the average S/H ratio 
in the Galaxy seems to be lower than in the Magellanic Clouds, contrary to the pattern of the remaining heavy 
elements. The  nitrogen abundances also follow the same pattern as O/H, Ar/H, and Ne/H. The N/H ratio is 
affected by the dredge-up episodes in the PN progenitor stars, especially the first dredge-up in low mass 
stars and the second dredge-up in intermediate mass stars, apart from hot bottom burning (HBB) in  the more 
massive progenitors. From our results, the average contamination from the PN progenitor stars is probably 
small. An example of the metallicity distribution is shown in Figure 1 for the Milky Way and  
the Magellanic Clouds, where we have considered Sample A and the IAG data, respectively. 

   \begin{figure}
   \includegraphics[width=.50\textwidth]{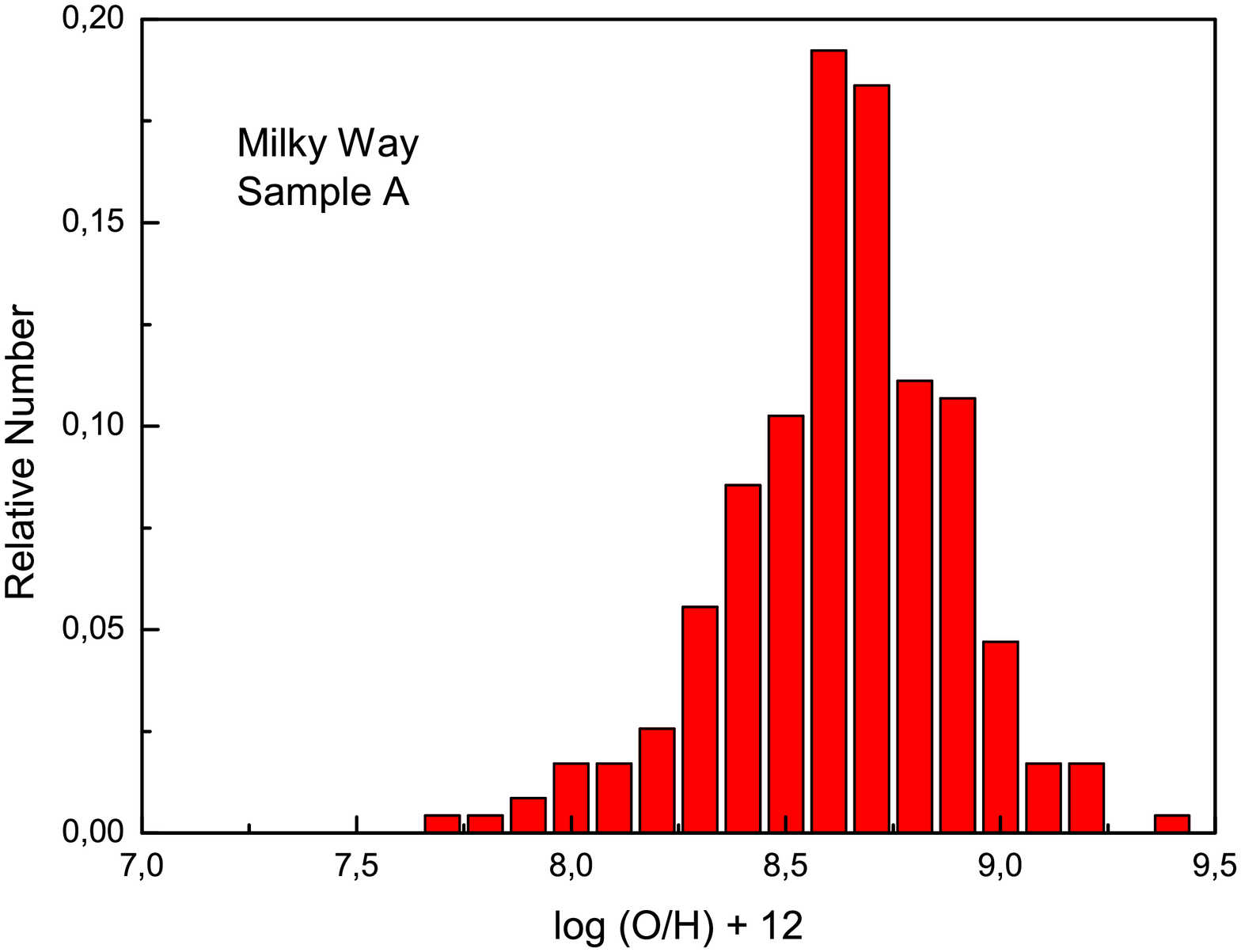}
   \includegraphics[width=.50\textwidth]{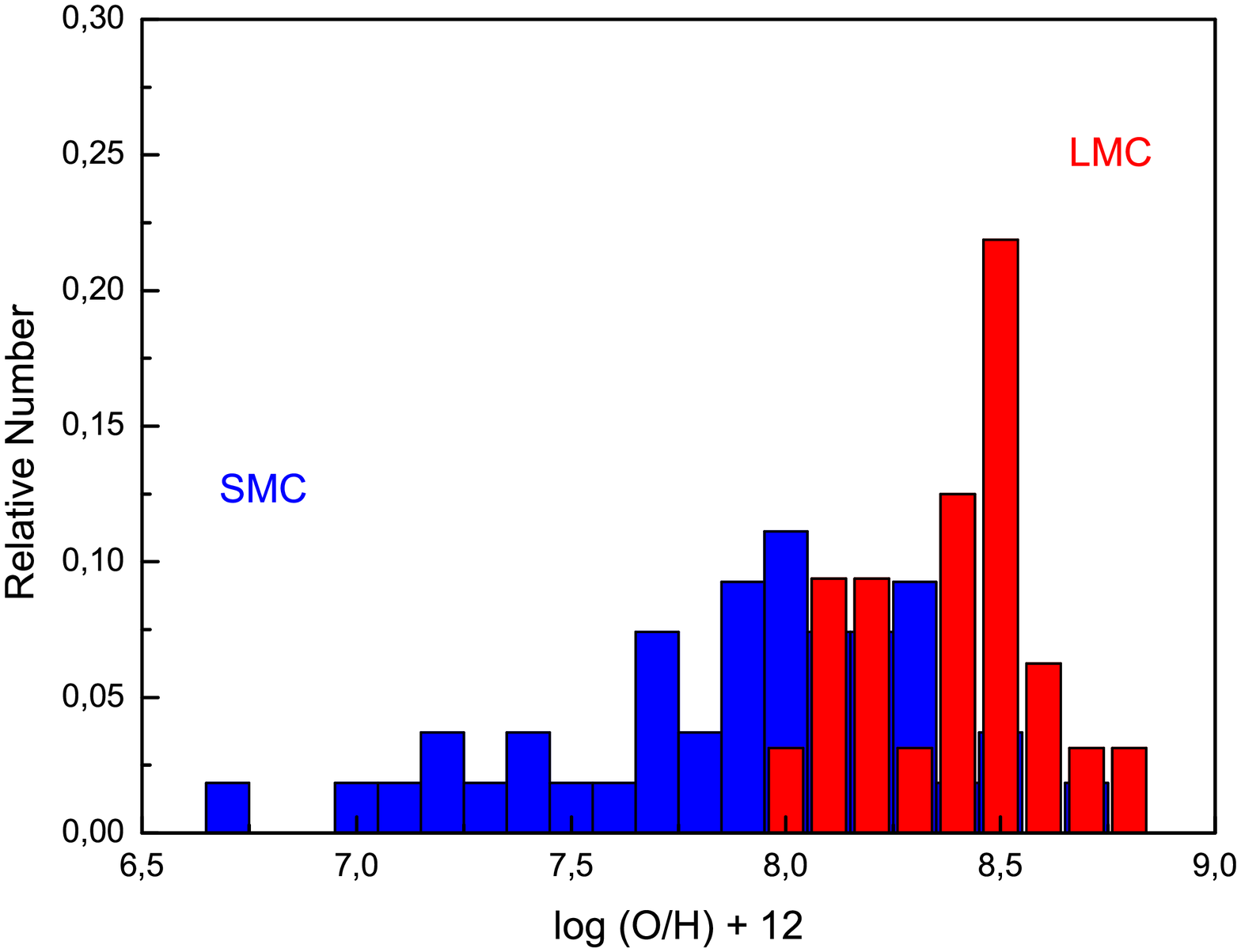}
   \caption{Average oxygen abundances of PN in the Milky Way (left) and Magellanic Clouds (right).}
   \label{histog}
   \end{figure}

\section{Abundance correlations}

The variation of the ratios Ne/H, Ar/H, and S/H with O/H usually shows a good positive correlation for all 
systems in the Local Group. This can be seen in Figure~2, where Ne abundances are plotted against the O/H 
ratio for the Milky Way and the Magellanic Clouds. Bulge PN, not shown in the figure, follow the same 
trend. The correlations are usually very good, with slopes in the range 0.8--1.2, with the exception of 
sulfur. As discussed in detail by Maciel et al. \cite{mci09}, there are still problems in the interpretation 
of sulfur data from PN, a conclusion supported by the comparison of the present results with some recent 
Spitzer data (Bernard-Salas et al. \cite{bernard}). It has been suggested (\cite{bernard}, \cite{henry04})
that the uncertainties in the S/H  abundances may be due to the possibility  that the adopted ionization 
correction factors overestimate the contribution of the S$^{+3}$ ion to the total sulfur abundances. 
Figure~2 also shows the Ne/O ratio, from which we note that this ratio is essentially constant for all 
metallicities with a relatively small dispersion, in contrast with  the elements produced by the progenitor 
stars, as we will see in the following.

   \begin{figure}
   \includegraphics[width=.5\textwidth]{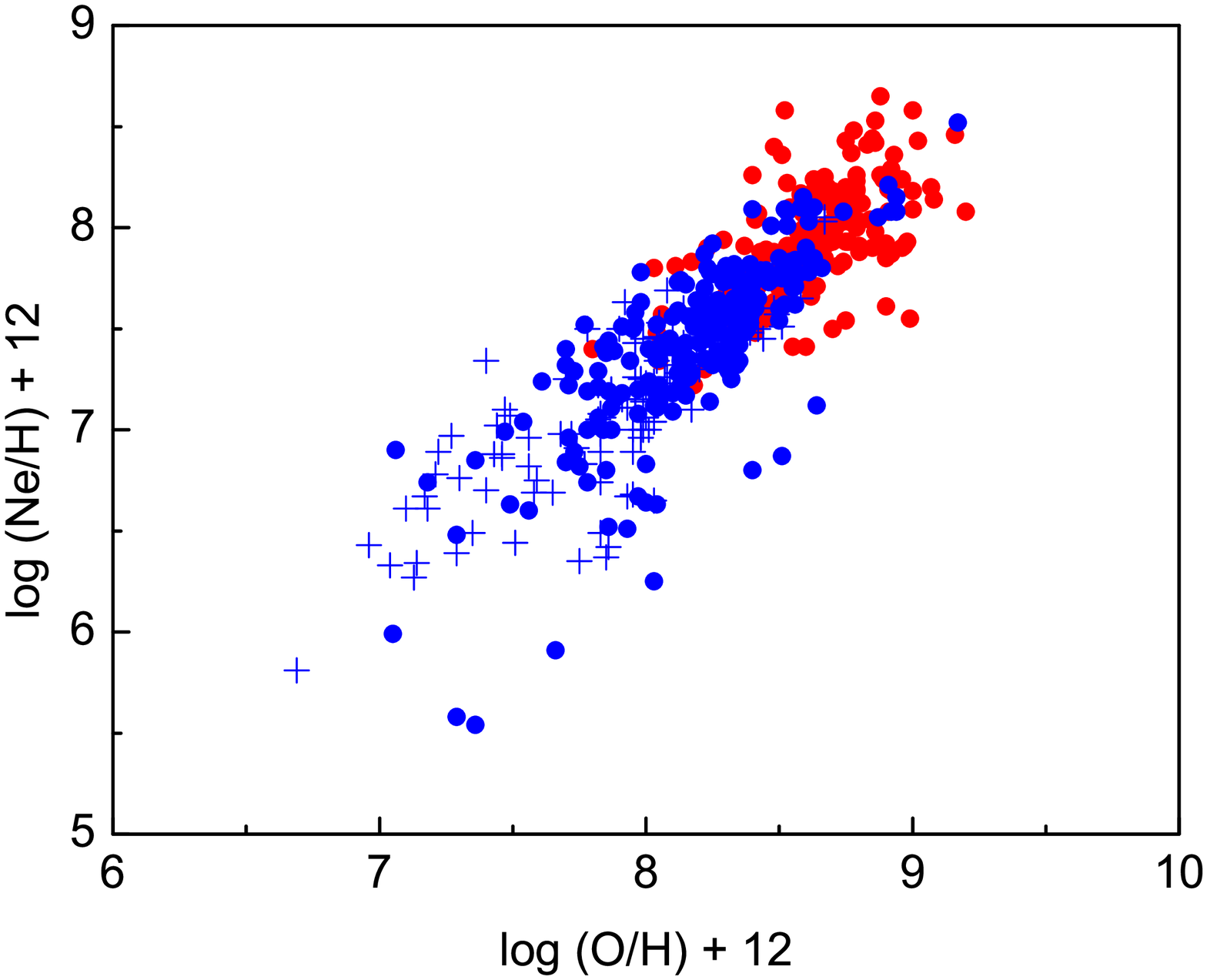}
   \includegraphics[width=.5\textwidth]{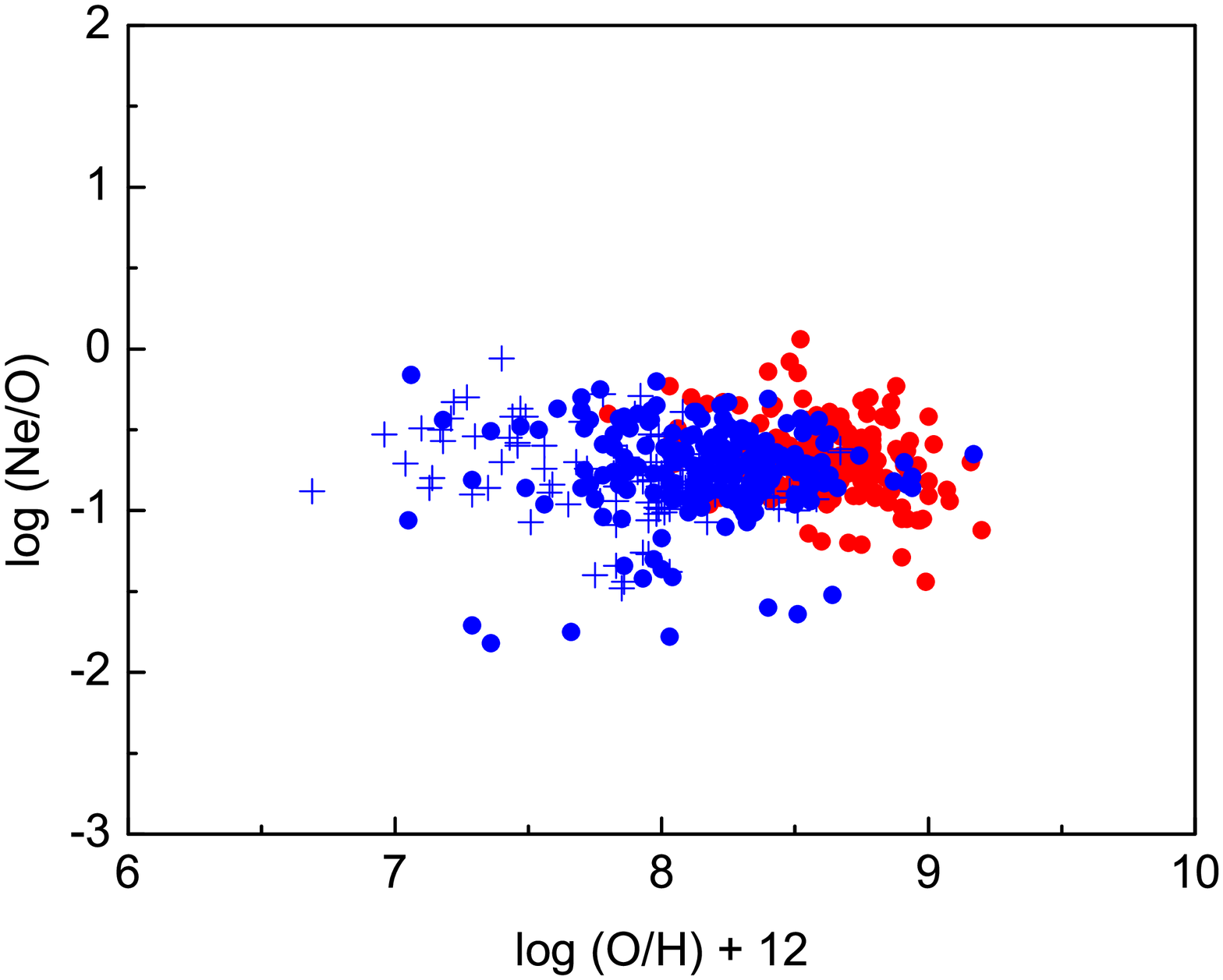}
   \caption{Neon abundances of PN in the Milky Way (red) and Magellanic Clouds (blue).}
   \label{neonio}
   \end{figure}

More interesting results can be obtained considering the elements that are produced during the evolution 
of the PN progenitor stars, namely, He, N, and  C.  A plot of the ratio N/H as a function of O/H 
leads to a positive correlation, as in the case of Ne and Ar, but  with a larger dispersion. This is due 
to the fact that the PN display both the original N present at the formation of the star plus the 
contamination by the progenitor star. In other words, the N/H ratio measured in PN shows some 
contamination, or enrichment, in comparison with the original abundances in the progenitor star.  The plot 
of the N/O ratio as a function of O/H  also shows a larger dispersion in comparison with the Ne/O plot 
shown in Figure~2, for the same reason. 

   \begin{figure}
   \includegraphics[width=.5\textwidth]{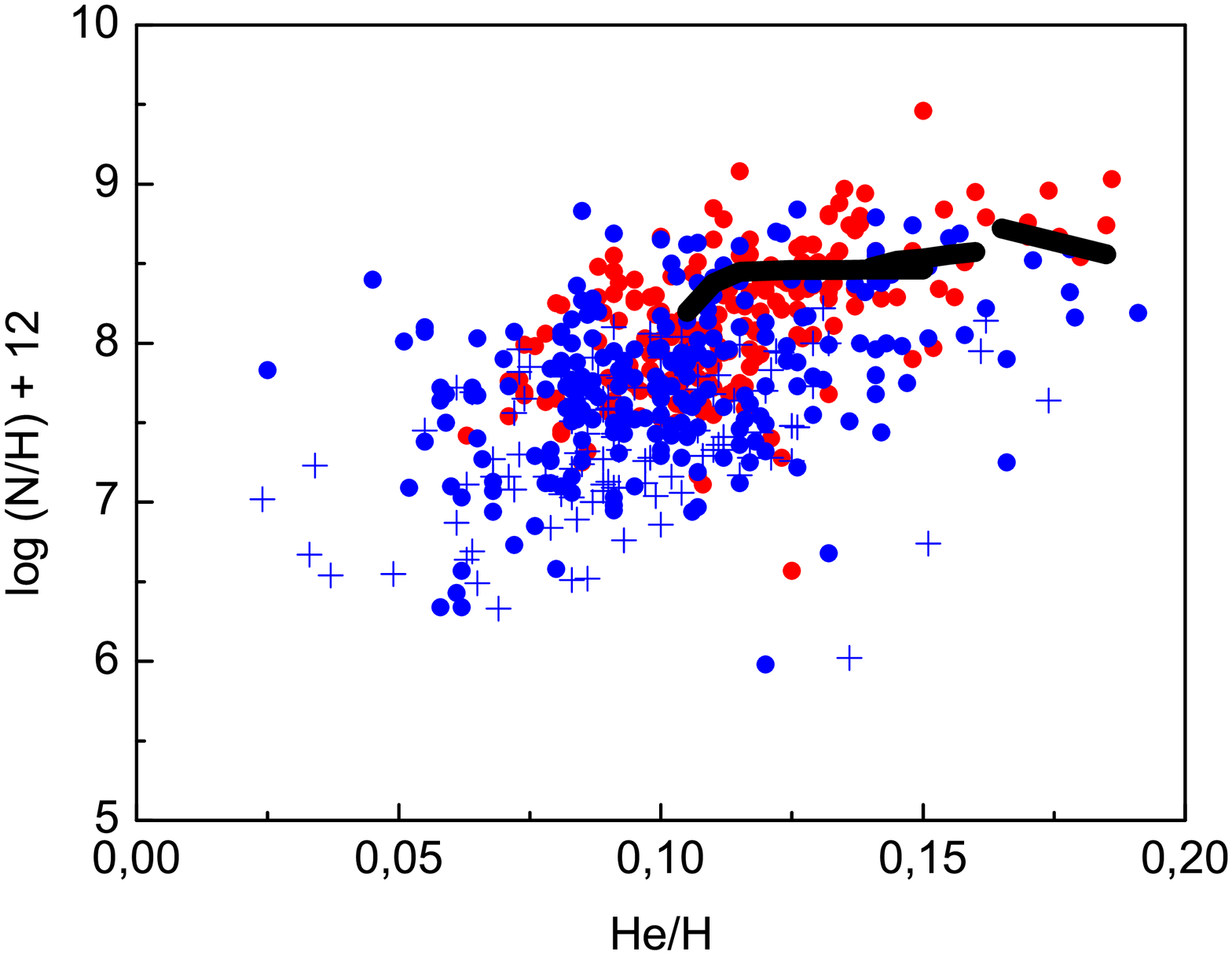}
   \includegraphics[width=.5\textwidth]{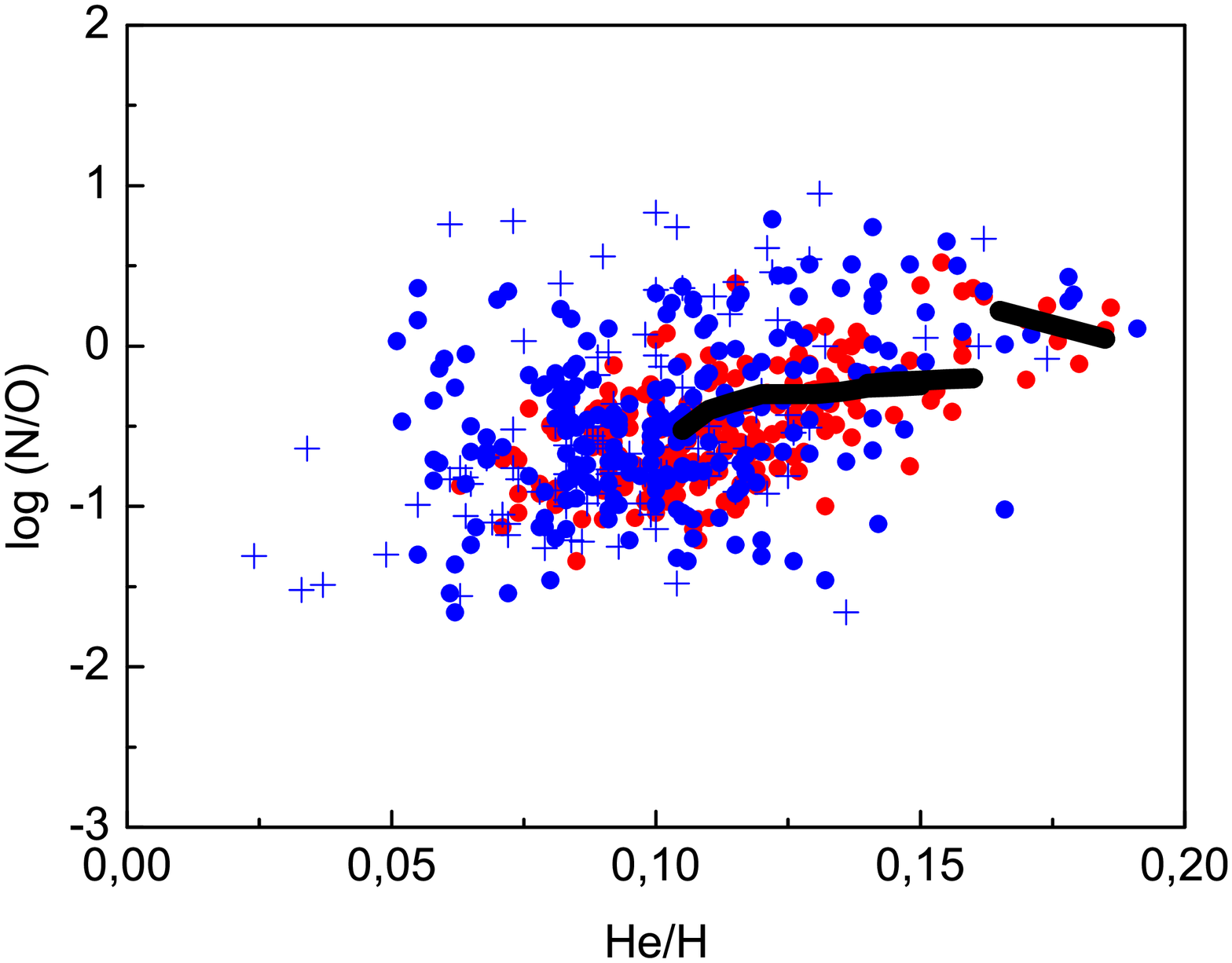}
   \caption{Nitrogen abundances of PN and comparison with theoretical models by Marigo.}
   \label{nhno-m}
   \end{figure}

Very interesting abundance correlations involving N abundances are those obtained as functions of the He 
abundances, as shown in Figures~3 and 4. It can be seen that  the evolution of the abundances is consistent 
with a similar trend for all systems, since the slopes are similar for all objects.  In figure~3 
we have also included the predictions of theoretical models by Marigo et al. \cite{marigo}, shown by the 
thick lines. These are synthetic evolutionary models for the thermally-pulsing Asymptotic Giant Branch 
stars (TP-AGB) with masses in the range 1.1 to 5 M$\,_\odot$, in which the first, second and third 
dredge-up episodes occur, apart from HBB for the most massive objects. These processes affect the He/H ratio, 
and in fact most objects present some enhancement compared to the solar values. According to Marigo et al. 
\cite{marigo}, progenitors having 0.9 to 4 M$\ _\odot$ and solar composition can explain the \lq\lq normal\rq\rq\ 
abundances, He/H $<$ 0.15, while for those objects with higher enhancements (He/H $>$ 0.15) larger masses 
are needed, in the range 4 to 5 M$\ _\odot$ plus an efficient HBB. Similar plots can be obtained using the 
models developed by Karakas (see for example Karakas and Lattanzio \cite{karakas}), in which case the initial 
masses range from 1.0 to 6.0 M$\ _\odot$, and metallicities $Z$ = 0.02, $Z$ = 0.008 and $Z$ = 0.004 have been 
considered, as shown in Figure~4.
   \begin{figure}
   \centering
   \includegraphics[width=.5\textwidth]{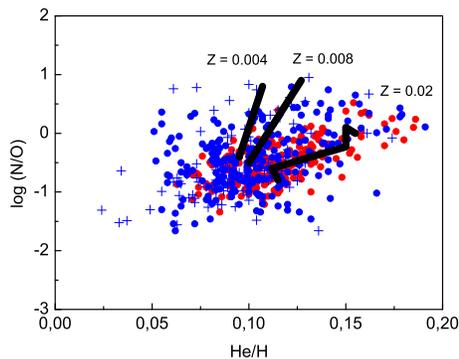}
   \caption{Nitrogen abundances of PN and comparison with theoretical models by Karakas.}
   \label{noheh-k}
   \end{figure}
The agreement for the N/O ratio is good, with a similar conclusion for N/H, and with a better agreement for 
the Milky Way, as expected, since this galaxy is somewhat more metal rich than the Magellanic Clouds.
We found a continuous transition between the \lq\lq normal\rq\rq\  and He-rich nebulae, which is probably 
due to the fact that we have a much larger sample compared to Marigo et al. \cite{marigo}. These results 
suggest that the nucleosynthetic processes occurring in these systems are similar, even though the global 
metallicity may be different and the chemical evolution may be affected by different star formation rates.
More detailed discussions can be found in some of our recent papers (\cite{thais}, \cite{mci09}, \cite{mci10}).


\begin{thebibliography}{99}

\bibitem{bernard} Bernard-Salas, J., Pottasch, S. R., Gutenkunst, S., Morris, P. W., \& Houck, J. R. 
    2008,  ApJ, 672, 274
\bibitem{henry04} Henry, R. B., Kwitter, K. B., Balick, B. 2004, AJ, 127, 2284
\bibitem{thais} Idiart, T. P., Maciel, W. J., Costa, R. D. D. 2007, A\&A 472, 101
\bibitem{karakas} Karakas, A., Lattanzio, J. C. 2007, PASA  24, 103
\bibitem{leisy} Leisy, P., Dennefeld, M. 2006, A\&A  456, 451
\bibitem{mc10} Maciel, W. J.,  Costa, R. D. D. 2010, Asymmetric Planetary Nebulae 5, (in press)
\bibitem{mci09} Maciel, W. J.,  Costa, R. D. D., Idiart, T. E. P. 2009, RMAA  45, 127
\bibitem{mci10} Maciel, W. J.,  Costa, R. D. D., Idiart, T. E. P. 2010, IAU Symp. 268, 
    ed. C. Charbonnel, M. Tosi, F. Primas, C. Chiappini
\bibitem{mcu} Maciel, W. J.,  Costa, R. D. D., Uchida, M. M. M. 2003, A\&A 397, 667
\bibitem{marigo} Marigo, P., Bernard-Salas, J.,  et al., 2003, A\&A 409,  619
\bibitem{stanghellini} Stanghellini, L. 2009, in IAU Symp. 256, The Magellanic Clouds: Stars, Gas, and Galaxies,
    ed. J. Th. van Loon \& J. M. Oliveira (Dordrecht: Kluwer), 421
\bibitem{stasinska} Stasi\'nska, G., Richer, M. G., \& McCall, M. 1998, A\&A, 336, 667

\end{thebibliography}
\end{document}